\documentclass[apl,twocolumn,showpacs,preprintnumbers,amsmath,amssymb]{revtex4}

\usepackage{graphicx}
\usepackage{dcolumn}
\usepackage{bm}
\usepackage[latin1]{inputenc}          
\usepackage[english]{babel}
\usepackage{color}
\usepackage{hyperref}

\begin{document}


\title{ On the magnetoelastic and magnetoelectric couplings across the antiferromagnetic transition in multiferroic BiFeO$_{3}$}  

\author{Mariusz Lejman$^{1}$, Charles Paillard$^{2}$, Vincent Juv\'e $^{1}$, Gwenaelle Vaudel$^{1}$, Nicolas Guiblin$^{3}$, Laurent Bellaiche $^{2}$, Michel Viret $^{4}$, Vitalyi E.  Gusev$^{5}$, Brahim Dkhil $^{3}$ and Pascal Ruello$^{1}$\footnote{ Electronic address: pascal.ruello@univ-lemans.fr}}

\affiliation{
$^{1}$Institut des Mol\'ecules et Mat\'eriaux du Mans, UMR CNRS 6283, Le Mans Universit\'e, 72085 Le Mans,  France\\
$^{2}$ Department of Physics and Institute for Nanosciences and Engineering, University of Arkansas, Fayetteville, United States.\\
$^{3}$Laboratoire Structures, Propri\'et\'es et  Mod\'elisation des Solides, CentraleSup\'elec, UMR CNRS 8580, Universit\'e Paris-Saclay, 91190 Gif-sur-Yvette, France\\
$^{4}$ SPEC UMR CEA/CNRS, L'Orme les Merisiers, France\\
$^{5}$Laboratoire d'Acoustique de Le Mans Universit\'e, UMR CNRS 6613, Le Mans Universit\'e, 72085 Le Mans, France}

\begin{abstract}

{Clear anomalies in the lattice thermal expansion (deviation from linear variation) and elastic properties (softening of the sound velocity) at the antiferromagnetic-to-paramagnetic transition 
 are observed in the prototypical multiferroic BiFeO$_3$ using a combination of picosecond acoustic pump-probe and high-temperature X-ray diffraction experiments. Similar anomalies are also evidenced using first-principles calculations supporting our experimental findings. Those calculations in addition to a simple Landau-like model we also developed allow to understand the elastic softening and lattice change at $T_N$ as a result of magnetostriction combined with electrostrictive and magnetoelectric couplings which renormalize the elastic constants of the high-temperature reference phase when the critical $T_N$ temperature is reached.}

\end{abstract}

\maketitle

\section{Introduction}

Multiferroic materials in which polarization, deformation and magnetic orders coexist have attracted continuous attention because of their tremendous potential in applications such as memories, spintronic devices, sensors/actuators or electro-optical systems as well as their underlying fascinating physics~\cite{wang,catalan}. Among them, the room-temperature multiferroic BiFeO$_3$ (BFO)~\cite{catalan} compound represents a rich playground system to understand the complex coupling mechanisms involving electric polarization, spin arrangement and phonon vibration allowing to manipulate their ferroic orders. Although the electrical control of the ferroelectric polarization and its domain distribution in BFO is nowadays well established, especially using piezo-force microscopy, manipulating the antiferromagnetic G-type order (AFM) which appears below N\'{e}el temperature $T_{N} \approx 650$~K, has revealed to be far more difficult, owing to the impossibility of a direct manipulation using magnetic fields. The existence of a magneto-electric (ME) effect, allowing the control of spins with electric fields, may just provide the long-sought route to control the AFM order. However, this ME effect remains weak because of Dzyaloshinskii-Moriya interactions known to be responsible for the non-collinear cycloidal modulation with 62~nm period~\cite{Sosnowksa1982,Rahmedov2012} which superimposes to the AFM order and prevents linear ME effect.

It is worth mentioning that in addition to electrical and magnetic stimuli, multiferroic properties can be most efficiently tuned using mechanical means. For instance, the ferroelectric polarization can be adjusted by using epitaxial strain or mechanical pressure~\cite{ingrid1, ingrid2,guennou,reviewYang,dup1,dup2,Edwards2018}. Similarly, the magnetic order can be also controlled with strain fields through the magneto-elastic coupling. As a result, the energy spectrum of thermally induced magnons (magnetic excitations) in BFO can be drastically modified with engineered static strain in thin films~\cite{strainmagnon,spinwave} or by pressure in bulk~\cite{Buhot2015}. Thus, both the magneto-elastic and magneto-electric couplings impact the magnetic arrangement in BFO, as recently shown in a neutron diffraction study~\cite{lee}.

Actually, the interplay between electric dipoles, magnetic orders and lattice dynamics is complex as further highlighted in a series of recent works. For example, thermal investigation of the phonon density of states (DOS) using neutron scattering revealed a coupling with magnons~\cite{neutron2}, and a softening of transverse acoustic branches in the vicinity of $T_{N}$ combined with a phonon peak broadening was as well observed near the AFM transition~\cite{neutron1}. This coupling between magnetism and strain in BFO is further confirmed with various techniques such as Raman scattering experiments~\cite{ramanmag,rovi}, or pulse echo method to evaluate the temperature dependence of the longitudinal elastic modulus ~\cite{smirnova}. This latter work showed a step-like form softening of the elastic constant ($\sim 0.7\%$) at the N\'{e}el transition. Yet, a theoretical work rather suggests a more monotonous change of the sound velocity in the vicinity of the N\'{e}el temperature without such softening~\cite{cao}. 

Despite this body of works, the mechanisms across the magnetic N\'{e}el temperature transition and the role of the various coupling interactions involved in BFO remain elusive. Here, using high-resolution X-ray diffraction (XRD) and picosecond acoustics pump-probe technique combined with first-principles calculations and Landau-based phenomenological approach, we attempt to unravel those mechanisms that couple the AFM order to the lattice dynamics by providing a global experimental and theoretical study of the lattice behavior through the N\'{e}el transition. We observe \textit{(i)} a significant and anisotropic change of the thermal expansion coefficients below $T_N$; and \textit{(ii)} a softening of the sound velocity of coherent transverse (TA) and longitudinal (LA) acoustic phonons below $T_N$, along with other related anomalies. As a matter of fact, significant effects on the lattice of BFO are revealed across the magnetic N\'{e}el temperature transition and discussed in the framework of magneto-electric and magneto-elastic couplings. Moreover, the existence of magneto-elastic changes at the GHz range open promising perspectives for the ultra-fast control of AFM (or magnetic) state by strain in future spintronic devices.

\begin{figure}[t!]
\centerline{\includegraphics[width=8cm]{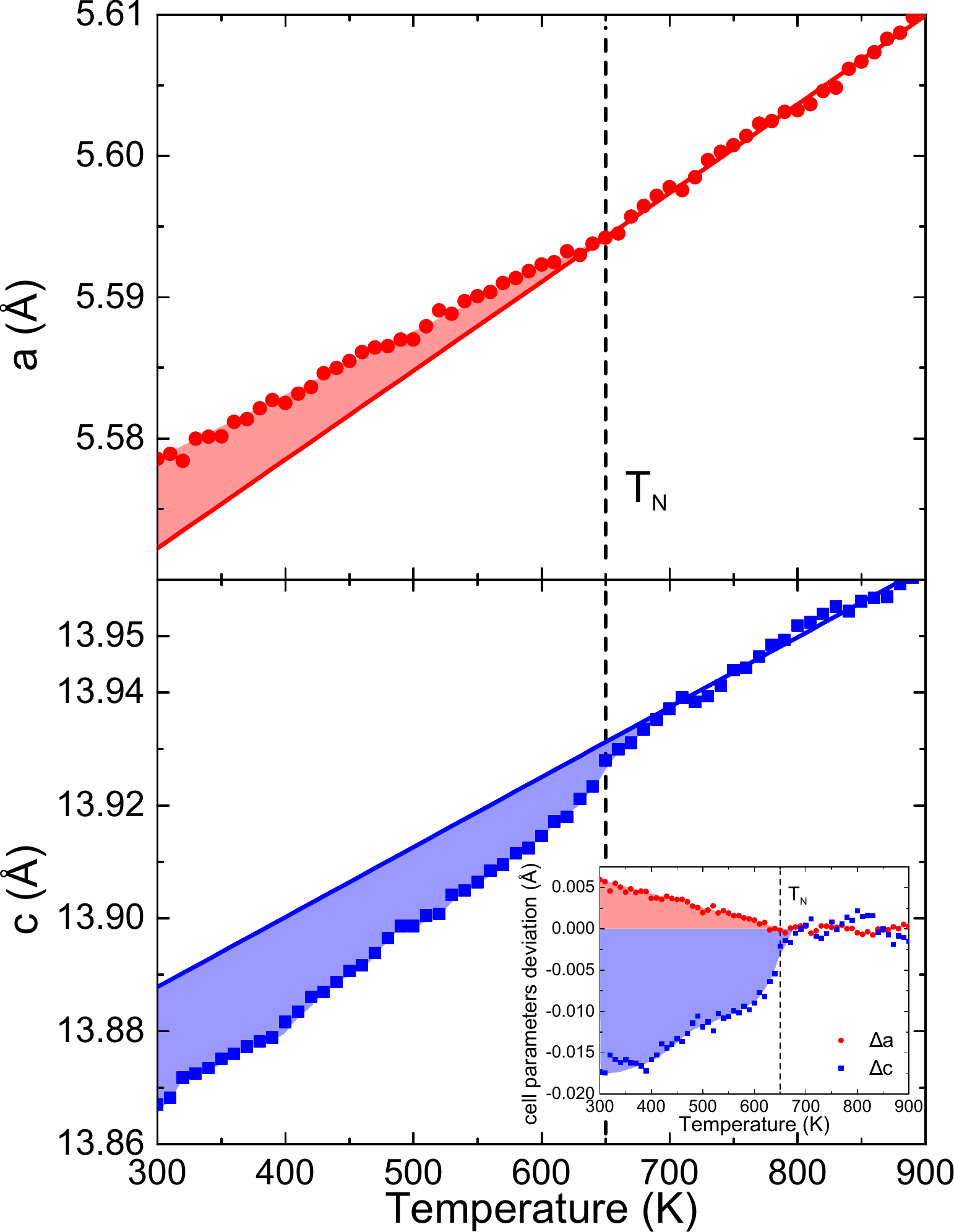}}
\caption{\label{fig1} (color online)  (a-b) Temperature dependence of the $a$ (top panel) and $c$ (bottom panel) lattice constants of the conventional hexagonal cell. Inset: difference of the unit cell parameters $a$ and $c$ arising below N\'{e}el temperature (vertical lines) from the deviation of the linear variation of the thermal expansion. }
\end{figure}

 \begin{figure}[t!]
\centerline{\includegraphics[width=8cm]{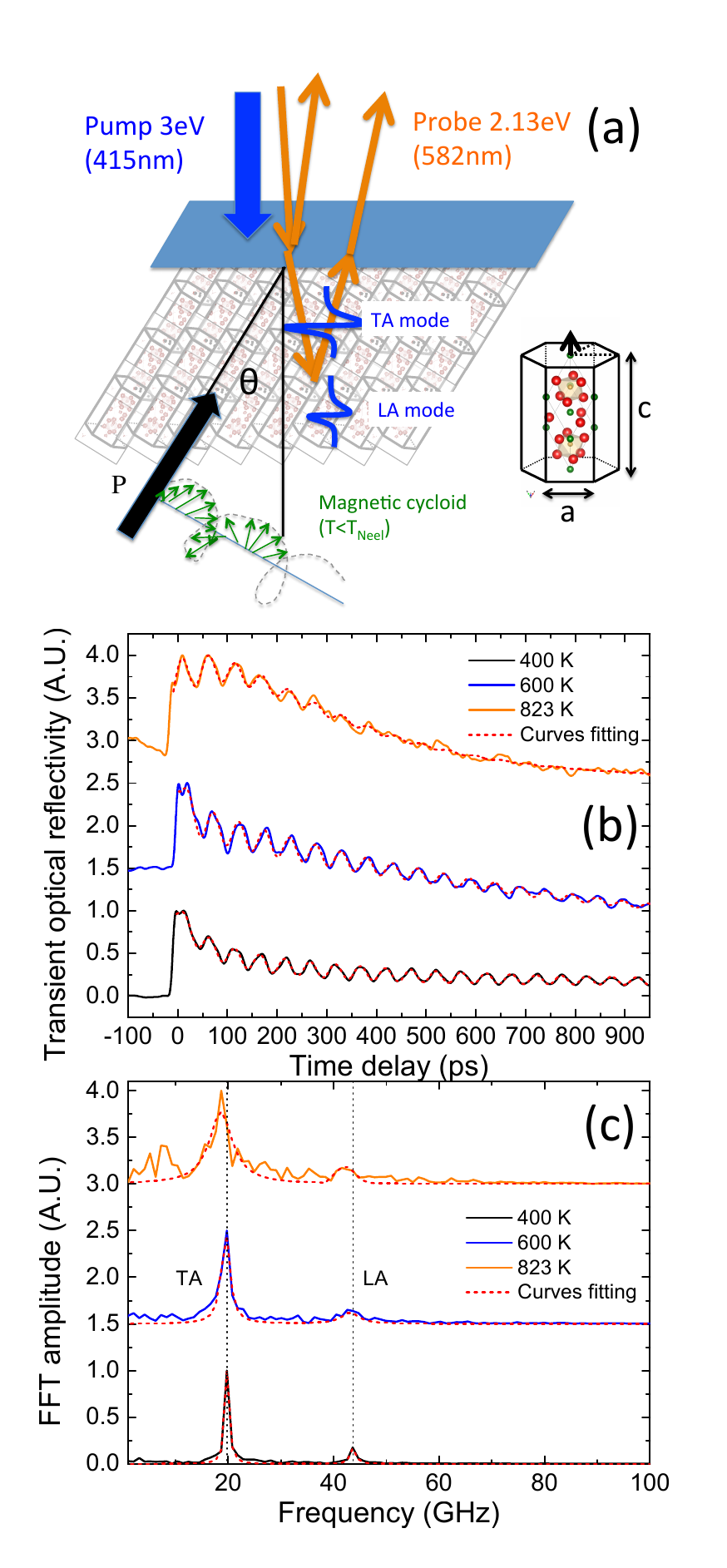}}
\caption{\label{fig2} (color online)  (a) sketch of the pump-probe experiment conducted with a disoriented BFO grain permitting to emit both LA and TA coherent acoustic phonons  \cite{ruelloapl,lejman2014}. (b) Temperature dependence of the time resolved optical reflectivity in BiFeO$_3$ for three selected temperatures. Note that the N\'{e}el temperature is about 650K. (c) corresponding Fast Fourier Transform (FFT) of the Brillouin components (TA and LA modes).}
\end{figure}

\section{Methods}

The sample under investigation is a polycrystalline BFO that was already described in previous works \cite{lejman2014,lejman2016}. 

The evolution of the $(104)_h$ and $(110)_h$ Bragg diffraction peaks (in the hexagonal conventional cell) with temperature ranging between 300K and 900~K (precision better than 1 K) was monitored using a high-resolution 2-axis diffractometer equipped with a rotating anode generator of 18 kW (Rigaku), with a Bragg-Brentano geometry and a 50 cm focalisation circle allowing an accuracy as high as 0.0002 $\mathrm{\AA}$  in 2$\theta$. The unit cell parameters $a$ and $c$ (the latter being along to the polar axis) in the hexagonal setting were then extracted (Fig. \ref{fig1}).

Coherent acoustic phonon dynamics were investigated with a pump-probe set-up developed for high temperature environment. It allows time-domain Brillouin scattering investigation with semi-opaque materials such as BFO \cite{ruelloapl,lejman2014,lejman2016}. Experiments were conducted on a specific grain of the polycrystalline sample, knowing each grain acts as a single crystal as detailed in Ref.~\cite{lejman2014}. The orientation of the polar axis with respect to the flat surface of the sample is roughly $\theta \sim 40^{\circ}$.
The BFO grain was photo-excited with above band-gap optical transition (2.6~eV is the bandgap of BFO~\cite{choi,nkT}) using a pump beam with 3~eV photon energy. The pump beam impinges on the sample surface with normal incidence, hence making a $40^{\circ}$ angle with respect to the polar axis of the grain, as shown in Fig. \ref{fig2}a. This pump beam generates LA and TA modes, as discussed in previous papers~\cite{ruelloapl,lejman2014}. The photogenerated coherent acoustic waves are then detected using a probe beam with 2.13~eV below bandgap energy, in normal incidence (the probe beams is inclined in Fig. \ref{fig2}a for clarity). The transient optical reflectivity of the probe beam $\Delta R (t)$ is monitored in time. The penetration depth of the probe beam ($> 1~\mu$m) allows to probe elasticity deep beneath the surface of the sample. Measurements were conducted in air. Heating and cooling cycles were performed to check for reproducibility of the results and stability of the sample. The highest temperature reached was 873~K \textit{i.e.} about 220~K above N\'{e}el temperature. Due to momentum conservation during the interaction between the probe light wave and the propagating acoustic front (for incident probe beam normal to the surface), only the acoustic phonon component with the Brillouin frequency ($f_{B}$) is detected~\cite{lejman2014,tom1}, accordingly to:

\begin{eqnarray}
\label{detect}
f_{B}=\frac{2nv_{s}}{\lambda}
\end{eqnarray}

where $\lambda$, $n$ and $v_{S}$ are the probe wavelength in vacuum, the refractive index of BFO at the wavelength $\lambda$ and the sound velocity in BFO (of either LA or TA waves), respectively. While BFO is birefringent~\cite{lejman2016}, the effect is limited in the considered orientation and only a mean refractive index was used in the present study. The typical transient optical reflectivity signal (Fig. \ref{fig2}(b)) is composed of a sharp increase when the material is excited by the pump laser followed by a decay in time  due to the electronic relaxation. Brillouin oscillations are superimposed on this baseline, as previously characterized~\cite{ruelloapl,lejman2014}. The Brillouin differential reflectivity signal $dR_{B}(t)$ can be cast in the general form:
\begin{eqnarray}
\label{ac}
dR_{B}(t) \propto sin(2\pi f_{B} t+\phi)e^{-\beta(T)\times t}e^{-\alpha(T) \times v_{s}(T) t}
\end{eqnarray}

where $\phi$ is the phase of the oscillatory signal and has both a contribution from the phonon field and from the optical detection process \cite{tom1}. $\beta(T)$ is the intrinsic anharmonic phonon term which governs the phonon damping. The partial penetration of the probe beam can also give rise to a damped signal as soon as the acoustic phonons leave the region of probe penetration. This contribution is described by the term $\alpha(T) v_s(T)$ where $\alpha$ is the optical absorption coefficient at the probe photon energy and $v_s$ is the sound velocity (LA or TA). Based on this formula and applying Fast Fourier Transform (FFT) (shown in Fig. \ref{fig2}(c)) or time-domain fitting, the frequencies of the TA and LA modes are extracted, as well as their damping time.

\begin{figure}[t!]
\centerline{\includegraphics[width=8cm]{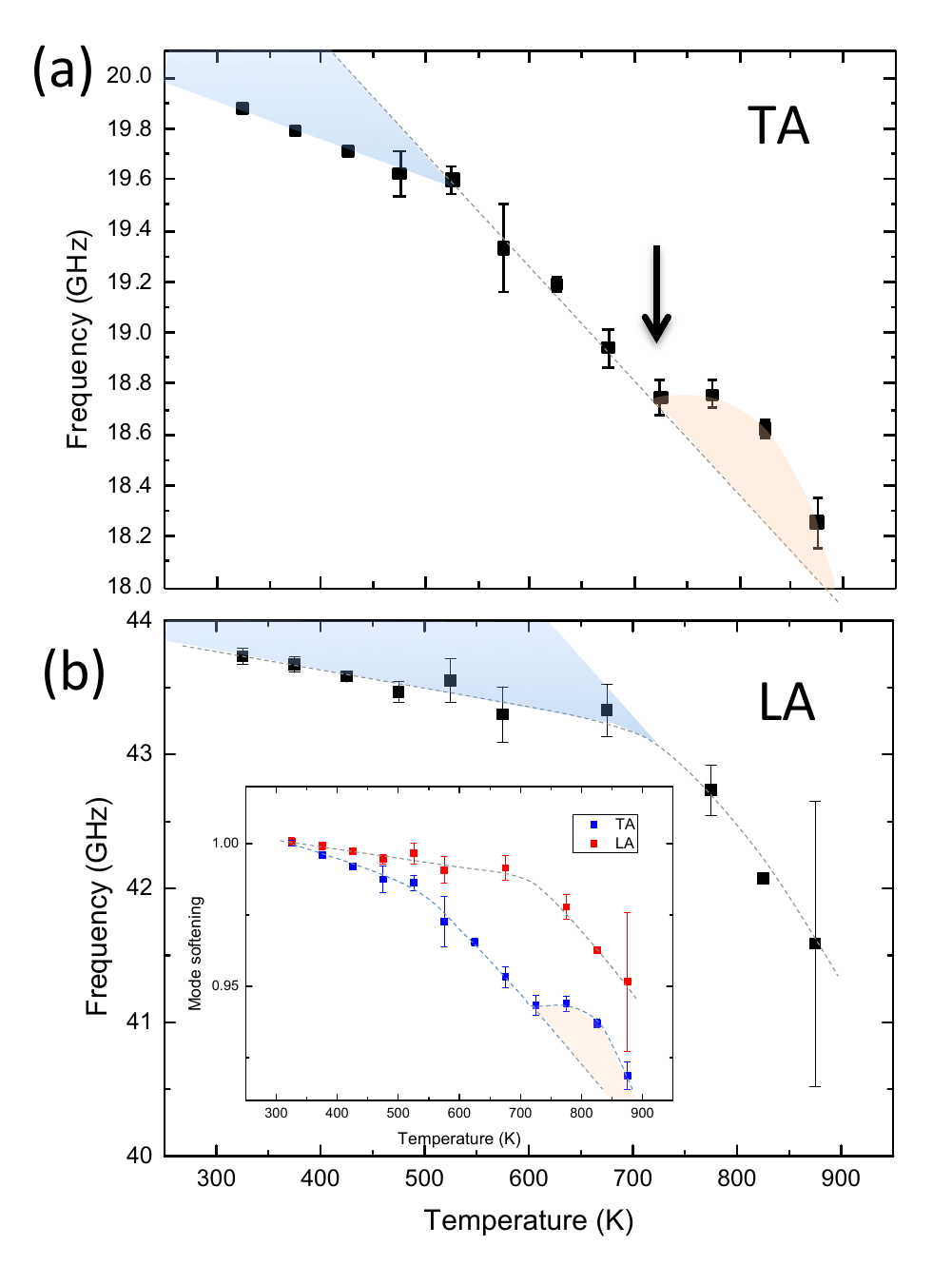}}
\caption{\label{fig3} (color online)  Temperature dependence of the Brillouin frequencies of the (a) TA and (b) LA modes. The inset in (b) shows the temperature evolution of normalized Brillouin frequencies.}
\end{figure}

In addition, Metropolis Monte Carlo simulations based on the effective Hamiltonian method described in Refs.~\citep{Kornev2006,Kornev2007,Albrecht2010,Prosandeev2013} were carried out. A $12\times 12 \times 12$ supercell of BFO, representing 8640 atoms, was considered. It was cooled down from 1500~K to 5~K under an applied electric field of magnitude $\sqrt{3}\times 10^9$ V.m$^{-1}$ applied along the $[111]$ pseudo-cubic direction to ensure that the low temperature phase is the monodomain $R3c$ ferroelectric ground state. Note that a G-type AFM order is obtained, and no cycloid order is considered in those calculations. $4\times 10^4$ MC sweeps were used during the field cooling procedure. Then, starting from the obtained low temperature field-cooled configuration, the field was removed and the supercell relaxed, starting at the temperature of 5~K and subsequently heating up. During this zero field heating phase, we used $10^6$ MC sweeps at each temperature. The various statistical averages used to extract relevant thermodynamic quantities were performed using the last $8\times 10^5$ MC sweeps.

\section{Results} 

Figures~\ref{fig1}(a,b) show the thermal evolution of the hexagonal lattice constants $a$ and $c$. Both unit cell parameters exhibit linear variation in the high-temperature region ($900~K>T > T_N$). Below the N\'{e}el temperature, there is a significant deviation for both lattice constants from the linear variation observed at high temperature, with a contraction of the polar axis $c$ and a change of the thermal expansion slope. This can be more clearly observed once the high-temperature trend is subtracted from the data, as shown in the inset of Fig.~\ref{fig1}b. The stretching of the polar axis length was previously observed in neutron diffraction experiments and attributed to the magnetostrictive-magnetoelectric coupling caused by the onset of the AFM order~\cite{lee}. In contrast, the lattice parameter $a$ exhibits a significant elongation (Fig. \ref{fig1}a) below $T_N$ along with a modification of the thermal expansion. This trend is also in agreement with Ref.~\cite{lee}. The overall volume of the hexagonal unit cell (shown in the Supplementary  Supplementary Figure 1 in Note I) is decreased in the AFM phase below $T_N$ with respect to the extrapolated volume from the high-temperature paramagnetic phase.

Interestingly, the kink observed at $T_N$ in the lattice constants, which are essentially static/time-averaged properties, are concomitant to anomalies observed in the lattice dynamics, as observed from the time-domain Brillouin signal. Therefore the AFM transition is also evidenced through acoustic waves (strain) at the GHz thus nanosecond time scale. We first note that the signal amplitude ratio between the TA and LA waves is not significantly affected at the AFM transition; as a result, the TA signal is always larger than the LA signal, as measured at room temperature (see Fig.~\ref{fig2}c). The temperature evolution of the TA and LA frequencies, obtained from the average value extracted from the FFT analysis and time domain fitting with Eq.~(\ref{ac}) is shown in Figs.~\ref{fig3}(a,b). It is clear that the Brillouin frequencies associated with both LA and TA modes do not have a linear dependence with temperature (see inset of Fig.~\ref{fig3}b). The TA wave shows a complex temperature behavior. It seems that the Brillouin frequency significantly softens in the vicinity of N\'eel transition ($710-850$~K) with respect to the high-temperature linear trend observed above 800~K, as shown by the arrow in Fig.~\ref{fig3}a. Below $T_N$ and more specifically below 710~K, the evolution of the TA Brillouin frequency can be described using two quasi-linear regimes with different slopes, the first occurring between 710~K and 570~K, and the second one below 570~K. The Brillouin frequency is connected to the speed of sound through Eq.~(\ref{detect}), and the speed of sound is related to the elastic constant $C$ as $v_s \propto \sqrt{\frac{C}{\rho}}$ with $\rho$ being the density of the medium. Hence the observed softening at $T_N$ may be explained by the softening of elastic constants, as for instance also observed by Smirnova \textit{et al.}~\cite{smirnova} in the longitudinal elastic constant. Nonetheless, direct comparison is impeded as Smirnova \textit{et al.} used a polycrystalline sample and thus sample an average of different crystallographic orientation (i.e., elastic constants) while here we study an individual grain/crystal. In contrast to the TA signal, no clear softening at $T_N$ is observed in the LA signal, possibly because of the large damping of the LA signal (see Figs.~\ref{fig2}(b,c)) which considerably affects the accuracy of the measurement. Nevertheless, the LA wave (see Fig.~\ref{fig3}b) clearly displays two linear regimes with different slopes, the change of regime occurring at the N\'eel temperature. 

Note that the anomalies observed at $T_N$ in the Brillouin frequencies of LA and TA modes could also come from anomalies in the optical index according to Eq.~(\ref{detect}). However, no significant change in the optical index was observed through the N\'{e}el temperature in previous reports~\cite{nkT}, and its evolution is thus considered to be linear with temperature. Besides, the global softening of the TA mode ($\sim 9\%$) is twice as large as that of the LA mode ($\sim 4.5\%$), thus indicating that the observed behaviors comes from elastic properties anomalies (see inset of in Fig.~\ref{fig3}(b)).  

We have also analyzed the damping time. Using the estimates of the absorption coefficient $\alpha(T)$ of Ref.~\cite{nkT}, and extrapolating up to 900~K, we have extracted the damping term   ($\alpha v_s$ term in Eq.~(\ref{ac})) coming from the finite probe light penetration length. We have found that this last term reproduces well the experimental observations, indicating it dominates over the intrinsic phonon damping characteristic time $\frac{1}{\beta(T)}$ in Eq.~(\ref{ac}) excluding the possibility to investigate the spin-phonon collision process.

\begin{figure}
\centering
\includegraphics[scale=0.7]{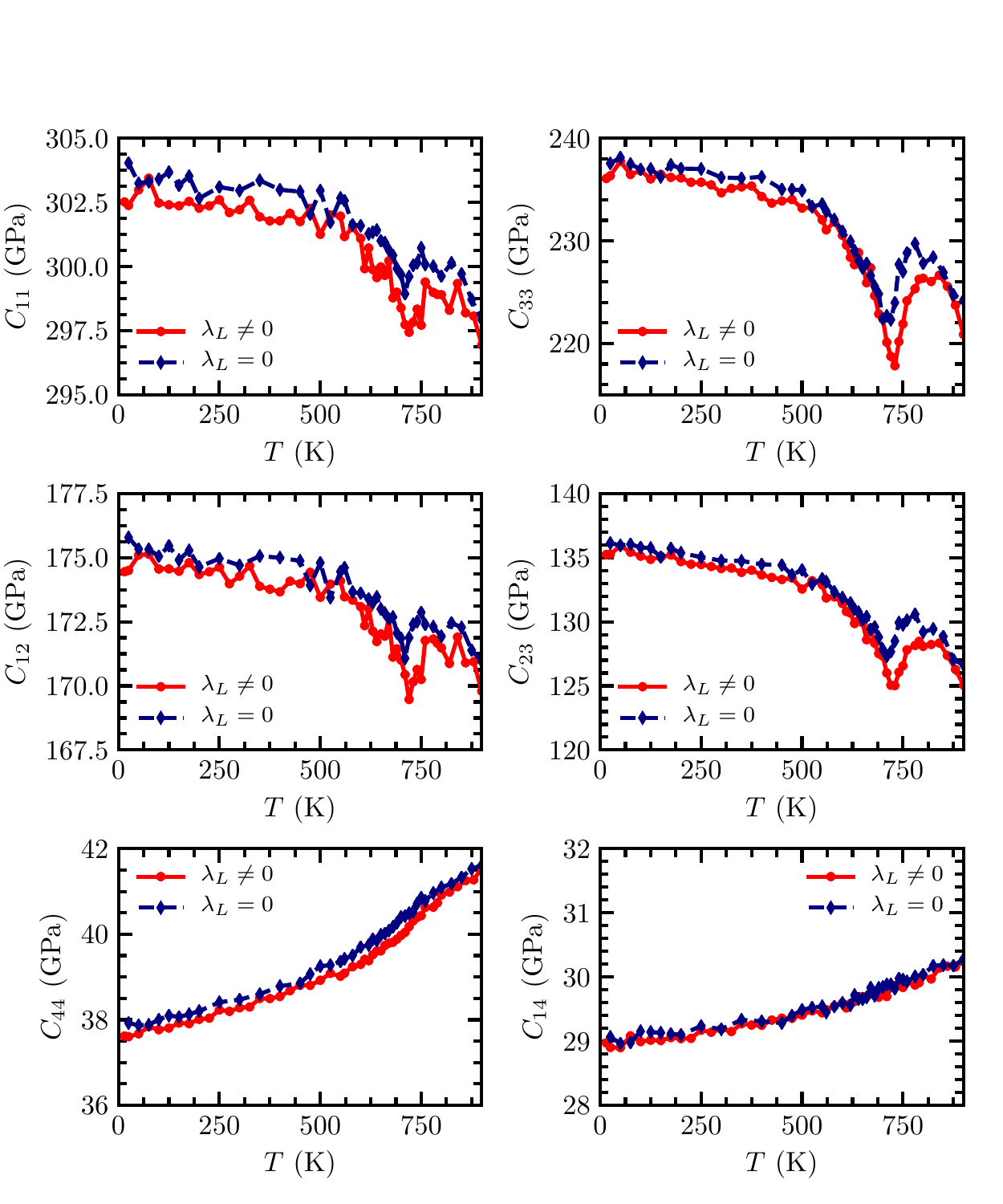}
\caption{Calculated independent elastic stiffness tensor components $C_{ij}$ in the conventional rhombohedral axes show anomalies at the N\'{e}el temperature (red curves). Results with the absence of the magneto-electric term ($\lambda_L=0$) are shown in blue curves.}
\label{fig4}
\end{figure}

\begin{figure}
\centering
\includegraphics[width=7cm]{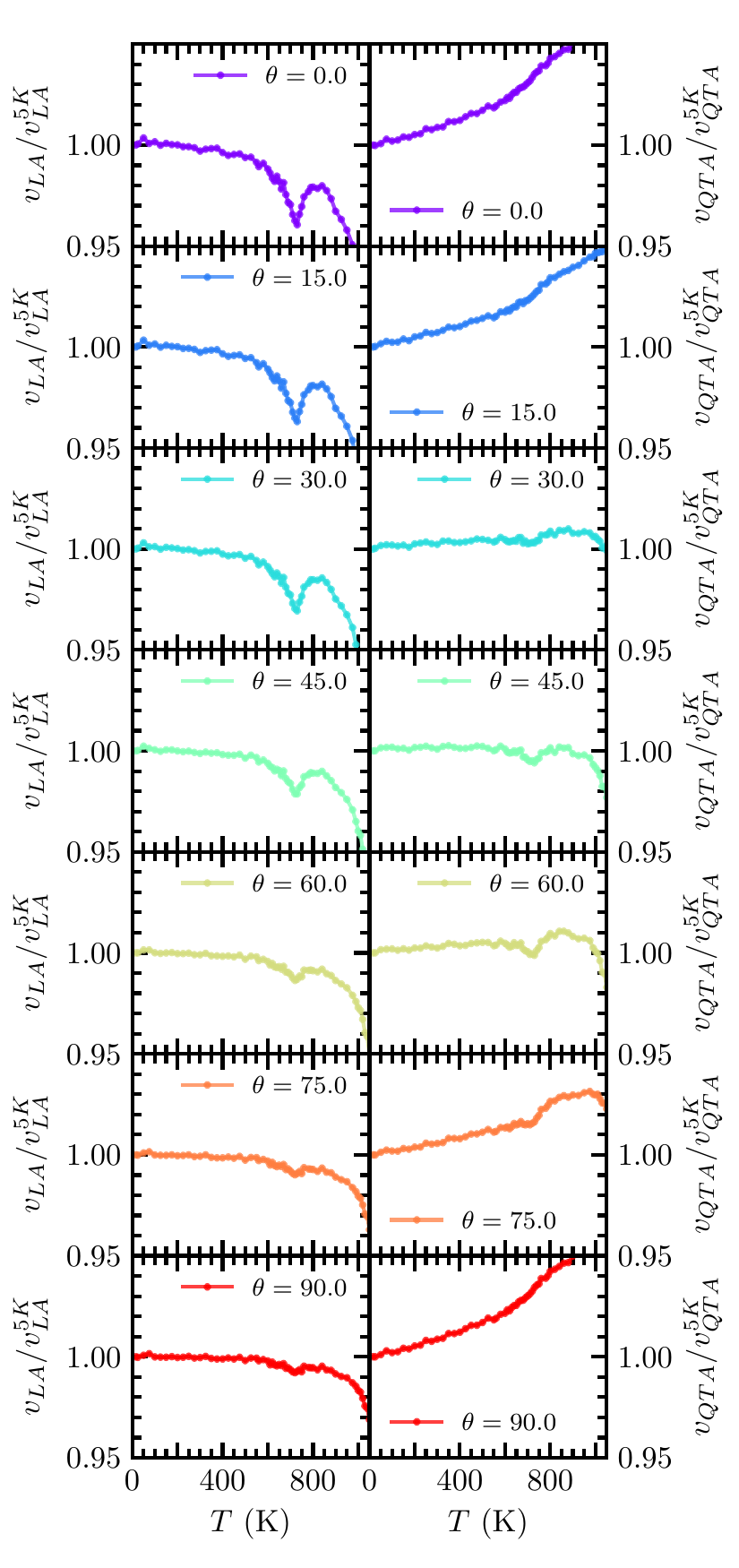}
\caption{Simulation of the temperature dependence of the LA (left columns) and TA (right columns) sound velocities for variable variable grain orientation angle $\theta$. The values are normalized to the value obtained at 5K.}
\label{fig5}
\end{figure}

In order to confirm the elastic origin of the anomalies observed in the LA and TA Brillouin frequencies, we used effective Hamiltonian calculations. The static elastic compliance tensor $S_{ij}$ in the pseudocubic axes was computed from the fluctuations of the homogeneous strain degrees of freedom,
 \begin{equation}
\label{eq:Sij}
S_{ij} = \frac{\left\langle V \right\rangle}{k_B T}\left(\left\langle \eta_i \eta_j \right\rangle - \left\langle \eta_i \right\rangle \left\langle \eta_j \right\rangle\right)
\end{equation}
with $\eta_i$ being a pseudocubic strain tensor component (in Voigt notation), $k_B$ being the Boltzmann constant, $T$ the temperature and $\left\langle V \right\rangle$ is the mean volume. Secondly,  the elastic compliance was transformed into the conventional rhombohedral axes system~\cite{Nye1985} (corresponding to the pseudocubic directions $[1\bar{1}0]$, $[11\bar{2}]$ and $[111]$ respectively), from which the elastic stiffness tensor $C_{ij}$ was derived. We depict them, in this latter frame, in Fig.~(\ref{fig4}). The elastic constants obtained at low temperature (5~K) agree well with previous calculations at 0~K from first-principles and related methods~\cite{Shang2009,Borissenko2013,Marton2017}. The obtained low-temperature evolution for $C_{11}$, $C_{33}$, $C_{12}$ and $C_{13}$ is rather flat until a significant softening is observed in the temperature range 550~K-850~K. In particular, a softening peak is observed at the computationally predicted N\'{e}el temperature of 730~K (see red curves Fig.~(\ref{fig4})). Further softening occurs above 850~K, caused by the proximity of the strong first-order ferroelectric-paraelectric phase transition which occurs at about 1100~K. We do not explore this region further as our focus is on the AFM-paramagnetic transition, and its implication on lattice properties. In addition, the transverse elastic constants $C_{44}$ and $C_{14}$ also exhibit a slight (but not as marked as $C_{11}$ or $C_{33}$ for instance) change above the N\'{e}el temperature, as the absolute value of their slope increases above $T_N$. The calculations were also carried out by "blocking" the direct coupling between spins and strain ($\lambda_L=0$, \textit{i.e.} no intrinsic magnetostrictive coupling) as shown by the blue curve in Fig.~(\ref{fig4}). We observe the same trend but with a slight attenuation of the softening at N\'eel temperature. In the next section, we discuss the possible origin of this elastic softening based on a simple Landau model including magnetostrictive and magneto-electric couplings.

\section{Discussion}

We have developed a simplified Landau model inspired from Refs. \cite{slon,carpenter} and describing \textit{(i)} the second-order phase transition with an antiferromagnetostriction term, \textit{(ii)} the first-order phase ferroelectric transition with an electrostriction term, and \textit{(iii)} a bi-quadratic magneto-electric coupling between polarization, $P$,  and antiferromagnetism. 
The antiferromagnetic order parameter $L$ is coupled to the strain $\eta$ through a linear-quadratic term $\frac{1}{2}\lambda_L\eta L^2$, and indirectly to it through the magneto-electric coupling term $\lambda_{LP} L^2 P^2$ combined with the electrostrictive term $\lambda_P \eta P^2$. Further details about the model are given in the Supplementary Note II. The thermodynamic potential $\Phi$ with respect to the high-symmetry cubic phase potential $\Phi_0$ can be cast as:

\begin{eqnarray}
\label{landau}
\Phi-\Phi_0 & = & {\color{red} \frac{1}{2}\alpha_L^0 (T-T_N) L^2+\beta_L L^2 } \nonumber \\
                  & + & {\color{blue} \frac{1}{2}\alpha_P^0 (T-T_0) P^2 + \frac{1}{4}\beta_P P^4 + \frac{1}{6} \gamma_P P^6}  \nonumber \\
& + & {\color{red} \frac{1}{2}\lambda_L \eta L^2} + {\color{blue} \frac{1}{2} \lambda_P \eta P^2} + {\color{green} \lambda_{LP} L^2 P^2} + \frac{1}{2}C^0 \eta^2,
\end{eqnarray}

with $ \alpha_L^0 > 0$, $\beta_L > 0$, $ \alpha_P^0 > 0$, $\beta_P < 0$ and $\gamma_P > 0$   constants; $C^0$ represents the elastic constants in the cubic phase, assumed to be temperature-independent. In other words, we neglect the anharmonic phonon coupling responsible for the temperature dependency of the elastic constants.
 
Based on the definition of the thermodynamic equilibrium, we can first derive the lattice strain induced in the AFM phase with respect to the strain caused by the ferroelectric order. It can be approximately cast in the form (see Supplementary Note II)

\begin{eqnarray}
\label{landaustrain}
\eta={\color{red} A \left( T_N - T \right)} + {\color{blue} B^{\prime} \left[  1 + \left( 1 + D (T_0 - T) \right)^{1/2} ~~~,\right]}.
\end{eqnarray}
where $A$, $B^{\prime}$ and $D$ are coefficients.

Secondly a renormalization of the elastic constant $C$ of the material in the antiferromagnetic phase ($T < T_N$) can be derived (see Supplementary Note II),


			\begin{eqnarray}
			C & = & C^0 {\color{blue} - \frac{\lambda_P^2}{2\beta_P} \frac{1}{1+\frac{2\gamma_P}{\beta_P} P^2}} {\color{red} - \frac{\lambda_L^2}{2\beta_L}} \nonumber \\
			   & - & {\color{green} \frac{\lambda_{LP}^2}{\beta_P \beta_L} \frac{1}{1-\frac{\lambda_{LP}^2}{\beta_P \beta_L} + \frac{2\gamma_P}{\beta_P}P^2} } \nonumber\\
			   & & \times \left[ {\color{blue} \frac{\lambda_P^2}{2\beta_P} \frac{1}{1+\frac{2\gamma_P}{\beta_P} P^2} } + {\color{red} \frac{\lambda_L^2}{2\beta_L}} - \frac{\lambda_L \lambda_P}{\lambda_{LP}} \right]. \label{eq:Ccoupled2}
			\end{eqnarray}

In Eq.~(\ref{eq:Ccoupled2}), one can clearly separate \textit{(i)} a softening induced by the magnetic transition in red, $- \frac{\lambda_L^2}{2\beta_L}$; \textit{(ii)} a softening at the ferroelectric transition, in blue; and \textit{(iii)} a mixing of the ferroelectric and magnetic softenings caused by the magneto-electric coupling $\lambda_{LP} L^2 P^2$ (in green). We note that the contribution \textit{(i)} was already mentioned in Ref.~\cite{smirnova}. The renormalization of the elastic constant is plotted in Supplementary Figures 2-4, where we discuss the different contributions (magnetostrictive, electrostrictive, and both the mechanisms at the same time). 
 
We can observe that the experimental strain renormalization of the lattice parameter $a$ (see inset of  Fig.~\ref{fig1}b) appears to be consistent with the linear term ${\color{red} A \left( T_N - T \right)} $ shown in Eq~(\ref{landaustrain}) as if the magnetostrictive term prevails in comparison to the electrostrictive-piezoelectric  term (i.e. ${\color{red}A}>{\color{blue}B^{\prime}} $). In contrast, the renormalization of experimental lattice parameter  $c$ (along which the ferroelectric order takes place) does not follow a linear temperature dependence (see inset of  Fig.~\ref{fig1}b) which may indicate in that case that both the magnetostrictive ($A$) and electrostrictive-piezoelectric ($B^{\prime}$) terms contribute, without being possible to provide at this stage a quantitative estimate. It is worth to mention that the tensorial nature of the Landau model can be accounted for and separate expressions for the $a$ (i.e., $\eta_1$) and the $c$ axis (i.e., $\eta_3$) can be obtained (see Supplementary informations), but the qualitative physical ingredients are already included in Eq.~(\ref{landaustrain}). 

In order to compare the experimental results of elastic properties with effective Hamiltonian calculations, we calculated the velocity of sound for different grain orientations, each of which involves different mixtures of elastic constants because of the inclination of the grain/crystal at an angle $\theta$ (see Supplementary Note III for details about the expression of $v_{LA}$ and $v_{TA}$). The determination of the TA and LA sound velocity has been performed with elastic constant computed with the magneto-elastic interaction terms ($\lambda_L \neq0$), see Fig.~\ref{fig4}. The results, plotted in Fig.~\ref{fig5}, show that the LA wave sound velocity should exhibit a clear softening peak at $T_N$ for all grain orientations. This theoretical prediction of the softening at the N\'eel temperature (dip) would be consistent with the observations of Smirnova \textit{et al.}~\cite{smirnova}. In our specific case ($\theta \approx 40^{\circ}$, close to $45^{\circ}$), this magnetic anomaly (dip of sound velocity) is not observed in Fig.~\ref{fig3}b, but the non-linear decrease of the Brillouin frequency with the temperature (Fig.~\ref{fig3}b), clearly different from a linear regime expected for a phonon-phonon anharmonic behavior, likely reproduces the envelop of the predicted renormalization of the elastic constant due to the magnetostrictive-electrostrictive coupling (see Supplementary Figures 1 and 2). The agreement with experiments is much better for the TA waves, since the calculated sound velocity value $v_{TA}$  for $\theta = 45^{\circ}$ in Fig.~\ref{fig5} predicts a small softening near $T_N$, which actually is consistent with the observed kink in the TA Brillouin frequency in Fig.~\ref{fig3}a (see arrow). However, calculations predict an almost temperature-independent TA sound velocity below 575-600~K, while a continuous variation of the TA Brillouin frequency is observed in Fig.~\ref{fig3}a. One should however remember that the effective Hamiltonian does not include the complete set of anharmonic phonon-phonon interactions, and thus the elastic constants of the high temperature cubic phase are nearly temperature independent, whilst they should typically harden at low temperature. As a result, we are missing a global linear baseline variation in the calculated sound velocity which may account for some of the discrepancies between our simulated results and the observed evolution of the Brillouin frequencies of the TA and LA waves.

The physical origin of these elastic anomalies in the vicinity of the N\'eel temperature is complex since a combination of magnetostrictive, electrostrictive and magnetoelectric effects actually exist as shown by the calculation based on the effective Hamiltonian. These couplings co-exist over the entire explored temperature range when we look at Fig.~\ref{fig4}.  In this figure we compare Monte-Carlo simulations performed with ($\lambda_L \neq 0$) and without ($\lambda_L = 0$) direct coupling of the magnetic moment with the strain in the effective Hamiltonian, in order to compare the relative strength of the direct magnetostrictive effect and the combined effect of magneto-electric and electrostrictive couplings. It can first be observed that the N\'eel critical temperature is slightly shifted down when the $\lambda_L$ coupling is disabled and, secondly, that the elastic anomaly is still present despite the absence of the direct coupling between magnetic moments and strain. Nonetheless, we can observe that the strength of the elastic anomaly (height of the pit) is reduced by 28\% and 38\% for $C_{11}$ and $C_{33}$ when the direct magneto-elastic coupling is turned off. As a result, it appears that the elastic anomaly at $T_N$ results from both magneto-elastic and magneto-electric effects.

\section{Conclusion} 

Using a combination of experimental and theoretical investigations of the complex coupling of polarization, magnetization and deformation in BiFeO$_3$, this work highlights the significant coupling of strain with the AFM order. Indeed, anomalies at the N\'{e}el temperature are detected both in experiments, via the evolution of Bragg peaks (static regime) and acoustic waves Brillouin frequencies (dynamic regime), and modeling \textit{via} the temperature evolution of the elastic constants using both first-principles and Laudau-based calculations. Our results indicate a predominant magnetostrictive contribution for the lattice parameter $a$ while a combination of magnetostrictive and electrostrictive effects are found to be active on the polar lattice parameter $c$. Interestingly, there is a qualitatively good agreement between the experimental observations (especially for the TA waves) and the calculated results in Figs.~\ref{fig3} and~\ref{fig4}. Additional picosecond acoustics experiments are now necessary to further explore the complex anisotropic effect our calculations have predicted for various crystal orientations as shown in Fig.~\ref{fig5}. Moreover, further pump-probe experiments could be also envisioned with probe wavelength in the near infra-red to increase the probe light penetration in order to extract  the damping of acoustic phonons coming from phonon-phonon and phonon-spin collisions. Besides these perspectives, the our results demonstrate that the existing coupling between magnetic ordering and strain paves the way for possible coherent control of antiferromagnons with coherent acoustic phonons, in line with the recent ultrafast magneto-acoustic experiments in ferromagnetic materials \cite{bigot,sherbakov, sacha, thevenard,tobey}.

\textbf{Acknowledgments}. M. L. would like to thank the Ecole Doctorale 3MPL (Matiere Molecules et Materiaux, Pays de la Loire France) for his PhD grant. C. P. and L. B. acknowledge the ARO grant W911NF-16-1-0227. V. J, G. V, V. G and P. R thank Le Mans Acoustique grant "Ferrotransducer project" 2015-2019. C. P. and B. D. thank a public grant overseen by the French National Research Agency 

\newpage

\end{document}